# Study of Compression Statistics and Prediction of Rate-Distortion Curves for Video Texture

Angeliki V. Katsenou[a,*], Mariana Afonso[a] and David Bull[a]

[a]*Office 1.23, Visual Information Lab, One Cathedral Square, BS1 5DD, Bristol, UK*

ABSTRACT

Encoding textural content remains a challenge for current standardised video codecs. It is therefore beneficial to understand video textures in terms of both their spatio-temporal characteristics and their encoding statistics in order to optimize encoding performance. In this paper, we analyse the spatio-temporal features and statistics of video textures, explore the rate-quality performance of different texture types and investigate models to mathematically describe them. For all considered theoretical models, we employ machine-learning regression to predict the rate-quality curves based solely on selected spatio-temporal features extracted from uncompressed content. All experiments were performed on homogeneous video textures to ensure validity of the observations. The results of the regression indicate that using an exponential model we can more accurately predict the expected rate-quality curve (with a mean Bjøntegaard Delta rate of .46% over the considered dataset), while maintaining a low relative complexity. This is expected to be adopted by in the loop processes for faster encoding decisions such as rate-distortion optimisation, adaptive quantization, partitioning, etc.

## 1. Introduction

Although recent video coding standards such as High Efficiency Video Coding (HEVC) (1), VP9 (2) and AV1 (3) have achieved impressive compression gains with significantly better rate-quality performance compared to their predecessors, they are all challenged by certain types of content, in particular complex dynamic textures (see examples in (4)). The Versatile Video Coding (VVC) (5) standard under development adopts a similar coding architecture and, while offering overall coding gains, still exhibits the same limitations. A recent statistical analysis of HEVC reference software (HM) performance has shown that the codec handles various types of texture very differently in terms of coding modes and bit rate (6). For example, for the homogeneous $256 \times 256$ video texture patches in the HomTex (7) dataset, HM requires, on average, twice the number of bits per pixel (bpp) for dynamic discrete textures (e.g. falling leaves) compared to dynamic continuous textures (e.g. flowing water) and five times higher than for static textures (e.g. a camera panning over grass). This work also showed that there is correlation between encoding statistics and texture types. For example, dynamic continuous textures tend to use more intra modes.

Knowledge about the compression characteristics of video content prior to encoding can be exploited in various situations, including: off-line rate-quality optimisation of video-on-demand streamed content, multi-pass encoding, rate control, statistical multiplexing, in loop rate-distortion optimisation (8; 9; 10; 11; 12). One way of obtaining such knowledge is through multi-pass encoding where encoder settings are adjusted according to post-encoding statistics (8; 11). Another approach is that embodied in the Netflix Dynamic Optimiser (9; 10). The algorithm consists of encoding video shots multiple times with different parameters (e.g. at different quantisation levels and/or different spatial resolutions); it then constructs a convex hull in rate-quality space and combines points from the convex hull to create an encode profile for the entire video sequence.

Other methods invest in using content features to build more efficient (13; 14; 15) or fast (16; 17; 18; 19) encoding mechanisms. In (13; 14; 15) the frames are segmented in textural areas and the depth decision/synthesis method/synthesis mode (respectively) is based on models built on the explored correlations of textural features to

---

All authors were with the Bristol Vision Institute and the Visual Information Lab, Department of Electrical and Electronic Engineering, University of Bristol, UK. The work presented was supported by the "Marie Skłodowska-Curie Actions- The EU Framework Programme for Research" project PROVISION, the Engineering and Physical Sciences Research Council (EPSRC), EP/M000885/1, and the Leverhulme Early Career Fellowship (ECF-2017-413).

*Corresponding author

✉ angeliki.katsenou@bristol.ac.uk (A.V. Katsenou)
ORCID(s): 0000-0003-0081-4488 (A.V. Katsenou)

compression or the content-tailored methodologies. In (16; 17; 18; 19) the authors explore the correlation of textural features to encoding statistics in order to make faster decisions for e.g. the prediction mode or the partitioning and, subsequently, reduce the encoding complexity.

For video streaming applications, that do not have the time constraints of a live transmission and have the advantage and flexibility of offline processing, approaches such as dynamic optimization have many advantages. However, this is not a generic coding solution, it is computationally intensive and comes with a financial cost (12). In such cases, reducing the computational overload would be beneficial. On the other hand, in cases where multi-pass methods are not appropriate, or where computing resources are limited, it would be helpful to have data extracted from uncompressed content that characterise the video's likely rate-quality performance in the context of the selected encoding method. Such a method could inform encoding decisions much more precisely and more efficiently than using multiple encoding techniques and statistics would only need to be extracted once from the original sequence.

The above challenges and observations provide the motivation to extract knowledge from uncompressed video content by analysing statistics of its spatio-temporal features with the aim of characterising the relationship between content and compression performance. Firstly, we advance on the works already presented (20; 6; 21) and further extend them, as explained next. In (21), we worked on identifying and defining the types of textures, recognizing features to work as their descriptors in the spatio-temporal or frequency domain and used those to classify them. Here, we provide more details of the feature statistics per texture type and we further demonstrate that such features correlate with the encoding statistics that were extracted and analysed in (6). We then explore the Rate-Distortion (RD) curves for the different texture types and show that it is beneficial to extract RDs per group of pictures. Based on this, we revisit the previously proposed mathematical model in (20) and develop suitable new mathematical models to represent these RD curves. Moreover, we explore the RD parameters relations to reduce the required predictions. For the training of the machine-learning based regression, we use a Random Forest (RF) based method to select the optimal feature set, which slightly varies for the different tested model parameters. Then, we perform a comparative analysis of different models on their prediction performance and their computational complexity. Concluding, we recommend a model that balances the tradeoff between the relative computational complexity and prediction performance.

The structure of the paper is as follows: Section 2 presents an analysis of the video texture, firstly as uncompressed content and secondly as compressed video through their statistics; Section 3 studies and models the RD curves of the homogeneous textures; Section 4 presents the results on the prediction of the RD curves for all considered mathematical models on homogeneous content. Finally, conclusions and future work are outlined in Section 5.

## 2. Analysing Video Content for Compression Purposes

The literature is rich in contributions relating to texture analysis (22; 23; 24) but most of these are in the context of computer vision-based recognition systems (25) and most deal only with still images. Our focus here however is the study of video compression performance. Video textures, particularly dynamic textures, are recognised as presenting significant challenges for video encoders (6). In this section, firstly we explore the relationships between various spatio-temporal features extracted from uncompressed videos in order to identify common characteristics. Secondly we encode a dataset of homogeneous video textures and independently study the associated encoding decisions and compression performance. Finally, we correlate the spatio-temporal features with the encoding decisions and performance. In order to perform meaningful video analysis, we have adopted the HomTex (7; 6) dataset; HomTex comprises homogeneous video annotated by experts. We also adopt the definition of video texture categories used in our previous work (21):

- Static: rigid texture that exhibits perspective motion, typically a moving solid object or a static background shot with camera motion, e.g. camera panning over a carpet.

- Dynamic Continuous: spatially irregular texture, with no clear structure, moving as a continuum e.g. water, deformable surfaces or smoke.

- Dynamic Discrete: spatially regular or irregular texture that consists of perspectively moving independent discernible parts or structures, e.g. leaves moving in a blowing wind.

The aforementioned categorisation have been validated in our previous work (6; 21) both from the perspective of the spatio-temporal characteristics of the uncompressed video content, as well as from the perspective of the encoding

Video Texture Compression

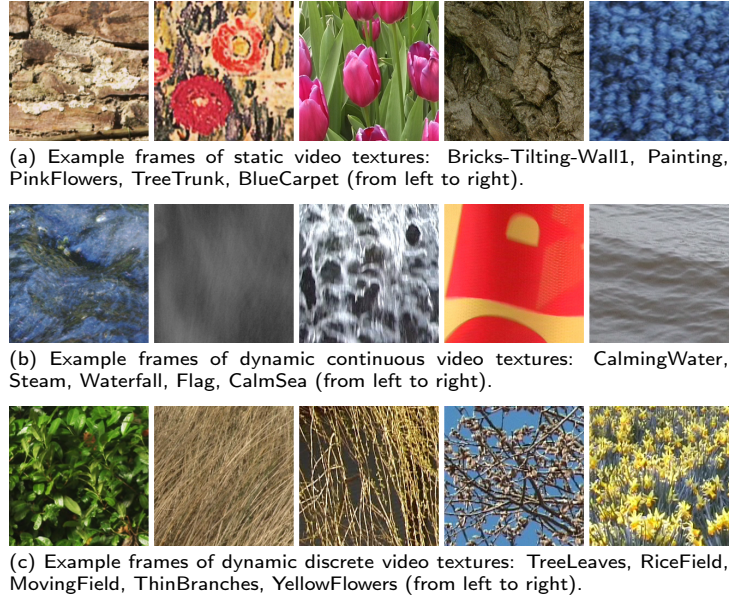

(a) Example frames of static video textures: Bricks-Tilting-Wall1, Painting, PinkFlowers, TreeTrunk, BlueCarpet (from left to right).

(b) Example frames of dynamic continuous video textures: CalmingWater, Steam, Waterfall, Flag, CalmSea (from left to right).

(c) Example frames of dynamic discrete video textures: TreeLeaves, RiceField, MovingField, ThinBranches, YellowFlowers (from left to right).

**Figure 1:** Examples of sequences from HomTex for the three different texture types that indicate the content variety.

**Table 1**
List of features and statistics (20; 21).

| Feature | Statistics |
|---|---|
| Gray Level Co-occurrence Matrix (GLCM) (23) | F1. $meanGLCM_{con}$, F2. $stdGLCM_{con}$, F3. $meanGLCM_{cor}$, F4. $stdGLCM_{cor}$, F5. $meanGLCM_{hom}$, F6. $stdGLCM_{hom}$, F7. $meanGLCM_{enr}$, F8. $stdGLCM_{enr}$, F9. $meanGLCM_{ent}$, F10. $stdGLCM_{ent}$ |
| Normalised Cross-Correlation (NCC) (26; 20) | F11. $NCC_{mean}$, F12. $NCC_{std}$, F13. $NCC_{skw}$, F14. $NCC_{kur}$, F15. $NCC_{ent}$ |
| Average Local Peak Distance (ALPD) (20) | F16. $ALPD_{mean}$, F17. $ALPD_{std}$ |
| Normalised Laplacian Pyramids (NLP) (27) | F18. $NLP_{mean}$, F19. $NLP_{std}$, F20. $NLP_{skw}$, F21. $NLP_{kur}$ |
| Temporal Coherence (TC) (20) | F22. $meanTC_{mean}$, F23. $stdTC_{mean}$, F24. $meanTC_{std}$, F25. $stdTC_{std}$, F26. $meanTC_{skw}$, F27. $stdTC_{skw}$, F28. $meanTC_{kur}$, F29. $stdTC_{kur}$, F30. $meanTC_{entr}$, F31. $stdTC_{entr}$ |
| Optical Flow (OF) (28) | F32. $meanOF_{mag}$, F33. $stdOF_{mag}$, F34. $meanOF_{or}$, F35. $stdOF_{or}$, F36. $meanOF_{curl}$, F37. $stdOF_{curl}$, F38. $meanOF_{ang}$, F39. $stdOF_{ang}$, F40. $stdOF_{covVx}$, F41. $meanOF_{covVy}$, F42. $stdOF_{covVy}$, F43. $meanOF_{covVxVy}$, F44. $stdOF_{covVxVy}$ |

decisions, statistics and performance. Examples of videos in these categories are illustrated in Fig. 1. For example, sequence Bricks-Tilting-Wall1 is a static texture where the camera pans over the wall. In the dynamic continuous texture CalmingWater, the camera is static but the water is moving continuously as it is triggered by a stream. In the dynamic discrete texture, TreeLeaves, a static camera captures the leaves moving irregularly in the wind.

### 2.1. Extracting Spatio-temporal Features and Statistics from Uncompressed Content

Textural features are conventionally defined with the purpose of facilitating similarity, browsing, retrieval and classification applications (22; 23; 24; 29; 30; 14; 15; 31). Additionally, most works have only considered static textures, namely images (23; 24; 22; 29; 30). Hence, most textural features do not capture the dynamic characteristics that texture exhibits in videos. Some of these features have previously been used for spatial segmentation in video synthesis and coding (13; 14; 15). An effort to synergise spatial and temporal texture features in video (but only for classification purposes) is reported in (31). Additionally, there are features in the literature that try to model the dynamic nature of texture, such as the motion co-occurence matrix (32) adopted also in (33).

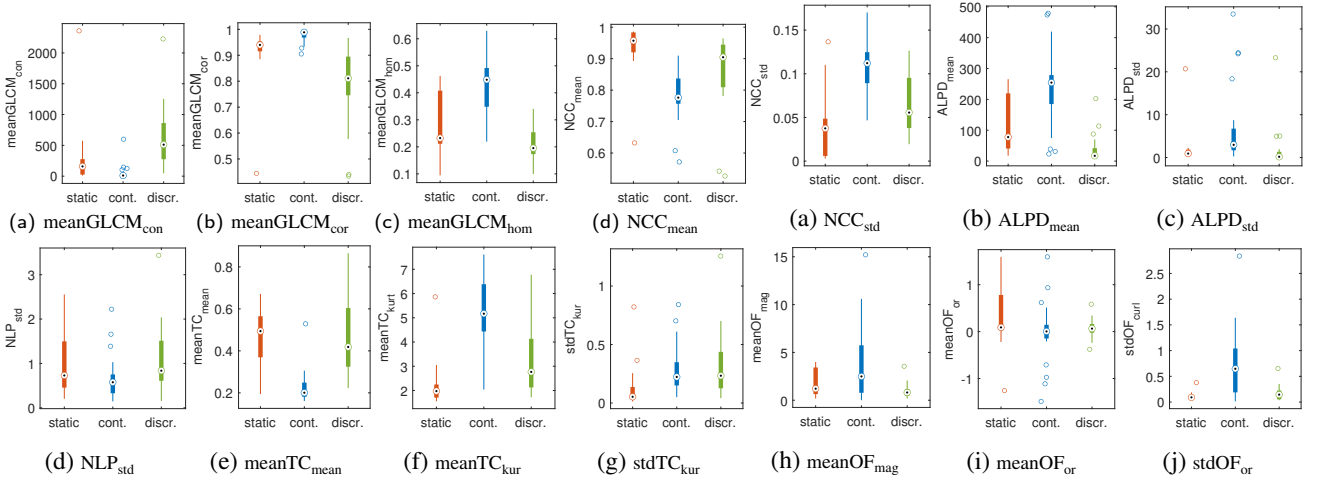

**Figure 2:** Distribution of several extracted spatio-temporal features and their statistics or the static (orange), dynamic continuous (blue) and dynamic discrete (green) video textures.

The spatio-temporal features employed in this work have been selected by investigating the vast variety of features in literature and by modifying some, so that they cover the basic characteristics of video texture that relate to encoding difficulty, i.e. spatial diversity, coarseness and motion, as shown in (21) (and as will be shown in Section 2.3). Note that these features are not a unique set and could be replaced by others that similarly capture the video texture characteristics. However, we adopt those that have been successfully used in our previous work for compression-related tasks, such as the recommendation of the perceptually-optimal frame rate (34) or for intelligently adapting spatial resolution prior to video encoding (35) or for the prediction of a content-driven bitrate ladder for adaptive video streaming (36).

Following the analysis used in the aforementioned works, six spatio-temporal features and their statistics (49 in total) were extracted as listed in Table 1. The GLCM (23) is a traditional spatial textural feature that expresses the intensity contrast of neighbouring pixels in an image, thus capturing the degree of coarseness and directionality of the texture. For the present frame $I_t$, let $G$ be the GLCM, whose element $G_{ij}$ is the number of occurrences for pixel pair $ij$ with intensity values $Y_i, Y_j$, with $Y \in \{0, 255\}$. The probability that a pixel pair $ij$ assumes $Y_i, Y_j$ values is $p_{ij} = G_{ij}/K$, where $K$ is the number of occurrences. GLCM has five main descriptors: contrast (con), correlation (cor), energy (enr) (or uniformity), homogeneity (hom) and entropy (ent) that are formally defined in the equations below:

$$\text{GLCM}_{\text{con}} = \sum_{i=1}^{M}\sum_{j=1}^{N}(i-j)^2 p_{ij}, \tag{1}$$

$$\text{GLCM}_{\text{cor}} = \sum_{i=1}^{M}\sum_{j=1}^{N}\frac{(i-m_r)(j-m_c)p_{ij}}{\sigma_r \sigma_c}, \tag{2}$$

$$\text{GLCM}_{\text{enr}} = \sum_{i=1}^{M}\sum_{j=1}^{N} p_{ij}^2, \tag{3}$$

$$\text{GLCM}_{\text{hom}} = \frac{p_{ij}}{1+|i-j|}, \tag{4}$$

$$\text{GLCM}_{\text{ent}} = -\sum_{i=1}^{M}\sum_{j=1}^{N} p_{ij} \log_2 p_{ij}, \tag{5}$$

where $M, N$ are the rows and columns dimensions respectively, $m_r, m_c$ the mean and $\sigma_r, \sigma_c$ the standard deviation

Video Texture Compressionalong rows and columns of both $I_t$ and $G$. We computed the mean values of the GLCM descriptors on a frame level and the mean and standard deviation of all these descriptors over the whole sequence, as seen in Table 1.

The NCC (26), which is commonly used in image processing applications for spatial similarity purposes, is used as in (20) as a spatio-temporal feature by capturing the peaks of cross-correlation between successive frames. It assumes values within the range $[-1, 1]$ with its maximum value indicating the maximum correlation and vice versa. In this paper, NCC is used as a spatio-temporal feature, as it examines the spatial similarity of two successive frames, $I_{t-1}$ and $I_t$, using a sliding matching template window $T$ of $w \times w$ size from the reference frame $I_{t-1}$:

$$\text{NCC} = \frac{\sum_{i=1}^{M}\sum_{j=1}^{N} |I_t(i,j) - \bar{I}_t(u,v)||T(i-u, i-v) - \bar{T}|)}{\sqrt{\left(\sum_{i=1}^{M}\sum_{j=1}^{N} |I_t(i,j) - \bar{I}_t(u,v)|\right)^2 \left(\sum_{i=1}^{M}\sum_{j=1}^{N} |T(i-u, i-v) - \bar{T}|\right)^2}} \qquad (6)$$

where $u, v$ define the area covered by the window $T$. By using an overlapping template between successive frames we computed the mean, the standard deviation, the skewness, the kurtosis and the entropy of the highest peaks between successive frames. All these statistics were averaged over the sequence length.

As an additional measure of the coarseness, we employed the ALPD in the third level of the Discrete Wavelet Transform (DWT) inspired by (22; 30). For all sequences, we computed the average local peak distance on a frame level, as below:

$$\text{ALPD} = \frac{1}{N}\sum_{n=1}^{N} \frac{\sum_{k=2}^{K} ||D_{3,p} - D_{3,p-1}||}{K}, \qquad (7)$$

where $k \in \{1, 2, \ldots, K\}$ the number of peaks and $D_{3,p}$ the peak $p$ in frame $n$ in the DWT domain. Then, we computed the mean and standard deviation over all frames.

If we assume that each frame $I_t$ is a distorted version of its previous neighbour $I_{t-1}$, we could use the NLPs (27) (or any other metric) to express this level of "distortion" at different scales as follows:

$$\text{NLP} = \frac{1}{N}\sum_{k=1}^{N} \frac{1}{\sqrt{N_s^{(k)}}} ||I_t - I_{t-1}||_2, \qquad (8)$$

where $N_s^{(k)}$ is the number of coefficients at scale $k \in \mathbb{N}$ and $N$ is the number of scales. Thus, NLPs attempt to capture hierarchical frame relationships. We computed the mean, the standard deviation, the skewness, the kurtosis and the entropy of the NLP between successive frames and then we averaged these statistics for the total number of frames.

In order to express how easy or difficult it is to predict one frame $I_t$ from its previous temporal neighbour $I_{t-1}$, we used TC, as in (20). It is computed using the Fast Fourier Transform (FFT) (37) and is defined as follows:

$$\text{TC} = \frac{|P_{I_{t-1}I_t}|^2}{P_{I_{t-1}I_{t-1}} P_{I_t I_t}}, \qquad (9)$$

where $P_{I_{t-1}}$ is the auto-spectral density of $I_{t-1}$ and $P_{I_{t-1}I_t}$ the cross-spectral density of frames $I_{t-1}, I_t$. TC is normalized within the range $[0, 1]$ and assumes its maximum value for static or purely translational motion among two successive frames. We computed the mean ratio of the square of auto-spectral density of the reference frame to the cross-spectral density of the reference frame to its successive frame, the standard deviation, the skewness, the kurtosis and the entropy between successive frames. Then, we took the mean and standard deviation of all these statistics for the length of the sequence.

The list of spatio-temporal features is completed with OF (28) that is based on a polynomial expansion. The OF descriptors and statistics are very important for the characterization of the dynamic textures, since the dynamic continuous textures exhibit different OF patterns compared to the dynamic discrete textures. We extracted the OF fields along with the following statistics: mean and standard deviation of magnitude, mean and standard deviation of orientation, mean and standard deviation of curl, mean and standard deviation of angular velocity, mean and standard deviation of covariance of horizontal OF vectors, mean and standard deviation of covariance of vertical OF vectors, mean and standard deviation of covariance of horizontal and vertical OF vectors.

Figure 2 depicts boxplots of the extracted features and their statistics, demonstrating feature distributions for the three types of video texture. Overall, many features demonstrate good selectivity between the different types of texture, notwithstanding some overlapping distributions. As expected, dynamic continuous textures express lower meanGLCM$_{con}$ values compared to static and dynamic discrete textures (see Fig. 2 (a)). This can be attributed to the lower density of edges (high spatial frequencies) in these types of textures. Due to this, the meanGLCM$_{cor}$ and meanGLCM$_{hom}$ (see Fig. 2 (b)-(c)) are higher compared to the other two types. We also note that the distributions of meanGLCM$_{con}$ and meanGLCM$_{cor}$ are wider for the dynamic discrete textures than for the static cases.

The mean and standard deviation of NCC (see Fig. 2 (d)-(e)) are, as expected, higher and lower, respectively, for the static textures compared to the dynamic types. This is because, for static video textures, the differences across successive frames are usually small. Regarding the dynamic textures, the mean NCC distribution ranges in higher values for discrete textures than continuous. The distribution of ALPD$_{mean/std}$ (see Fig. 2 (f)-(g)) is narrow and in a low value range for dynamic discrete textures as anticipated. This is attributed to the fact that these video textures exhibit dense edges that subsequently leads to a low peak distance. On the contrary, ALPD$_{mean/std}$ is higher for dynamic continuous textures and for the same reasons NLP$_{std}$ (see Fig. 2 (h)) are lower compared to the other two types. A good way to discriminate dynamic continuous texture from static and discrete is by the meanTC$_{con/kur}$ values (see Fig. 2 (i)-(j)), as it is significantly lower/higher, respectively. This is attributed to the low density in edges and the deformable nature of these type of content, which results in lower cross-spectral density. Lastly, although the OF statistics (see Fig. 2 (l)-(n)) appear less selective compared to the other features, they show a clear difference in the span of the magnitude and orientation of the OF vectors. Particularly, for dynamic discrete textures that are characterised by very fine local motion of usually fine grained content, the distribution is narrow and in the lower range of values. On the other hand, meanOF$_{mag}$ is wider for static and even more for dynamic continuous textures. These two can be well differentiated by the mean/stdOF$_{or}$ that are significantly different, namely the OF orientation has a higher but quite uniform value for static textures, while it is smaller on average but more variant across frames for the dynamic continuous textures.

## 2.2. Analysing Compressed Video Content by Extracting HM Encoding Statistics

HEVC encoding statistics were extracted using the test model version HM16.20 and the encoding analyser software, Harp (38). All the sequences from the HomTex dataset were encoded using the Main profile and three configurations: Random Access, Low Delay and All Intra. The initial quantization parameter (QP) was set to five commonly used values QP={22, 25, 27, 32, 37} that capture the rate-distortion curve. A total of 39 statistics were obtained from the encoding process at the Coding Tree Unit (CTU) level. These were then post-processed to obtain the statistics from the encoding decisions and performance per sequence for the different frame types (I, B and P). Table 2 summarises the encoding decisions and statistics that were extracted. For the measure of correlation between the original and the residual frames, the 2-D Pearson product-moment correlation coefficient was used, considering only the luminance component.

Figure 3 depicts the distributions of a subset of the statistics for the B frames of the Random Access configuration, using a QP value of 25 for all three types of texture (as annotated by experts).

The prediction modes selected per Coding Unit (CU) vary significantly for different texture types (see Fig. 3 (a)-(d)). As expected, static textures are associated with a high percentage of Skip mode, similarly low percentages of Merge and Inter and almost zero usage of Intra mode due to the simplicity of the motion present (camera panning or zooming). Dynamic continuous textures rely mostly on Intra mode, with fewer Inter, Skip or Merge modes. This implies that motion prediction frequently fails for these textures. Discrete dynamic textures, on the other hand, rely mostly on Inter and Merge mode as they exhibit distinct motion that can be effectively predicted, with reduced use of Intra mode.

Since different texture types exhibit different spatial patterns, it is expected that the number of partitions in a CTU will also differ. In particular, when the content has fine texture (high granularity), then the CTUs are usually highly divided. Thus, as seen in Fig. 3 (e), the highest average number of partitions per CTU is observed for discrete dynamic textures (with a median equal to 35), then for continuous dynamic textures (median of nine), and the lowest number is recorded for static textures (with a median equal to 4).

The RD performance also varies with texture type, as demonstrated in Fig. 3 (f)-(g). As expected, static textures require a smaller number of bits (a median value of 0.0043 bit per pixel (bpp)) to encode compared to dynamic textures that require 30 times more bpp for dynamic continuous and 80 times higher bpp for the dynamic discrete case, while exhibiting lower distortion (SAD of residual) for the same QP. It is also interesting to observe that, although dynamic discrete textures require on average 1.8 times higher bitrate compared to continuous, they also increase distortion by, on average, 1.97 times.

**Table 2**
Statistics extracted from HM during the encoding process (6).

| Category | Statistics |
| --- | --- |
| Prediction modes | intra (%), stdIntra, Skip (%), stdSkip, merge (%), stdMerge, inter (%), stdInter |
| Reference indexes | ref0 (%), ref1 (%), ref2 (%), ref3 (%) |
| Partitioning | avgPart, stdPart |
| Bits | avgBits, stdBits |
| Distortion | avgDist, stdDist |
| Bit allocation | bitsModeSignal (%), bitsPart (%), bitsIntraDir (%), bitsMergeIdx (%), bitsMotionPred (%), bitsResidual (%), bitsOthers (%) |
| Residual Statistics | avgMSEresi, stdMSEresi, avgMSERecError, stdMSERecError, avgCorrResi, stdCorrResi, avgCorrCodedResi, stdCorrCodedResi |
| Intra mode | DCIntra, PlanarIntra, avgIntraDir, stdIntraDir |
| Motion Vectors | avgLengthMV, stdDistMV |

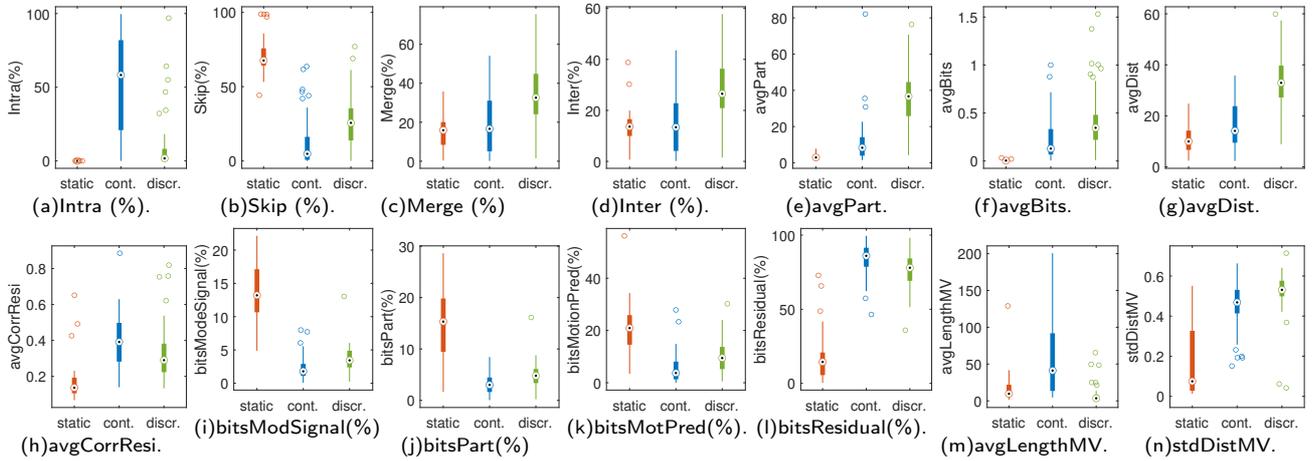

**Figure 3:** Distribution of a subset of the extracted encoding statistics for static (red), dynamic continuous (blue) and dynamic discrete (green) video textures.

The bit allocation for both types of dynamic textures in Fig. 3 (l) shows that the majority of the total bits are spent on residual coding, 77% on average of the bits generated for dynamic continuous and 84% for dynamic discrete video texture. Taking also into account that the other bit statistics used for coding additional information such as motion vectors, mode signalling, etc. (see Fig. 3 (i)-(k)) are only contributing a small percentage to the total encoding bitrate. We can also see that the residual for dynamic textures typically exhibits very high energy and this explains the high RD statistics. This is further confirmed by the high distortion and high correlation between the original frame and the residual as illustrated in Fig. 3 (h). Contrary to dynamic textures, static textures exhibit a more uniform bit allocation among the different categories as well as a low correlation of the residual signal to the reference.

Motion is another important characteristic. Static textures are associated with small magnitude motion vectors that show directional consistency. As expected, dynamic discrete textures have generally small magnitude motion vectors with high values in the distribution of directional irregularity. On the other hand, continuous textures are associated with a wide range of magnitude motion vectors with a slightly lower median value of irregularity in the direction.

## 2.3. From Textural Features to HM statistics

In this subsection, we validate the the use of spatio-temporal features from uncompressed content to predict encoding behaviour and performance. Figure 4 depicts a visualisation of the linear correlation matrix computed among the spatio-temporal features and the encoding decisions and statistics. The colorbar indicates the range of values, where blue colors indicate positive correlation and red colors indicate negative correlation. Higher linear correlation is represented by darker blue or red colours (for values close to 1 or −1, respectively.) This matrix can be seen as a way of finding the "strongest" candidate features in order to build a prediction models for the encoding performance using spatio-temporal features.

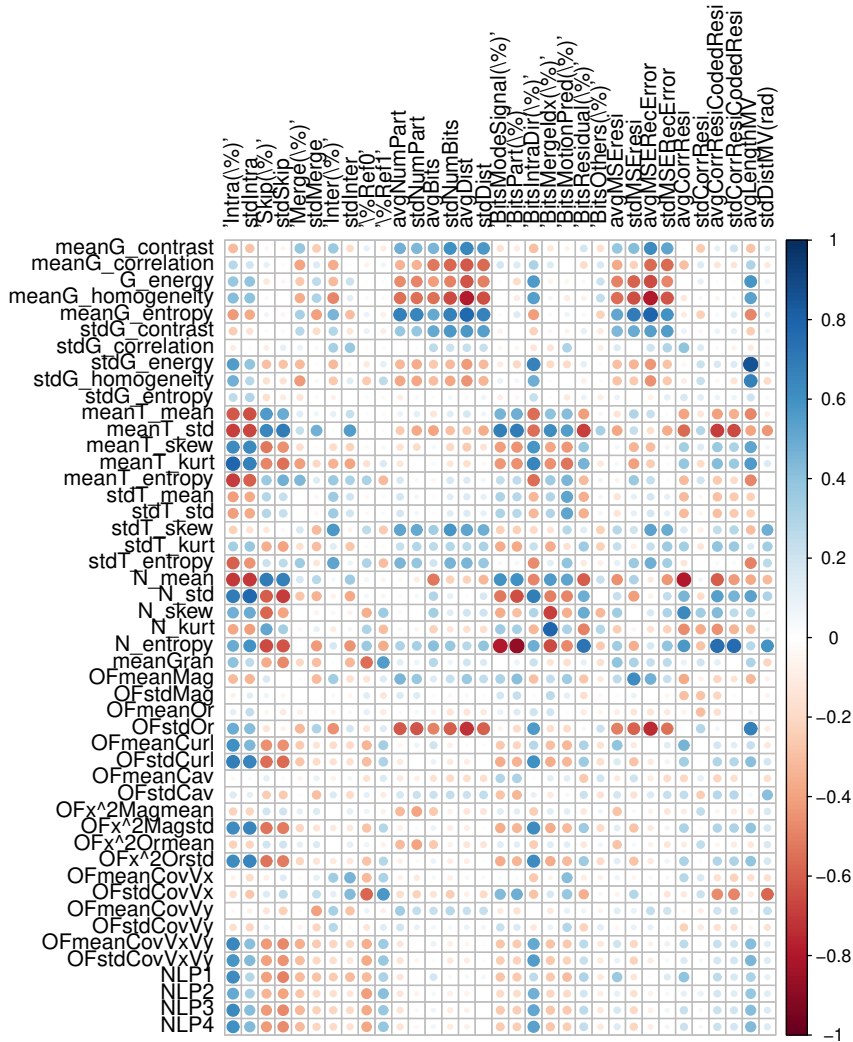

**Figure 4:** Pearson correlation matrix of the extracted spatio-temporal features and the encoding decisions and statistics.

As can be seen from the the correlation matrix, not all of the features demonstrate strong correlation to all of the encoding decisions and statistics. For some of the features and their statistics, however, the correlation matrix shows a strong linear relationship, indicating that the spatio-temporal features could be utilised to predict encoding performance. Particularly, the GLCM descriptor statistics are highly correlated with the partitioning statistics (avg/stdPart), the number of bits required for encoding (avg/stdBits), the resulting quality (expressed either as avg/stdDist or as avg/stdMSERecError) and the residual statistics (avg/stdMSEresi). The TC statistics are highly correlated to certain prediction modes, i.e. Intra(%), Skip(%) and Inter(%), as can be seen in Fig. 4. The prediction modes are also highly correlated to NCC statistics. Furthermore, the NCC statistics are correlated to the allocated bits statistics and the resid-ual signal reconstruction error. Compared to the aforementioned features, the OF and NLP statistics show lower linear correlation values to the encoding decisions and statistics. The highest correlation values for OF reported in the matrix concern the standard deviation of the OF vectors' orientation of the partitioning statistics and the reconstruction error.

## 3. Predicting RD Curves

In this Section, we explore the links between RD curves and video texture characteristics, we study their behaviour for different texture classes and fit them using mathematical models. One of the few approaches to understanding the coding performance of video textures was in (39), where the authors categorise sequences as static, dynamic and mixed and study the RD curves of the compressed sequences using HEVC and and the Advanced Video Coding (AVC) (40) reference software. To the best of our knowledge, there exists no other reported work that studies the RD properties of homogeneous video textures.

### 3.1. Comparison of Mean and per GoP RD Curves

For the purposes of our study, we require a large amount of homogeneous video texture data. We extract the RD curves per Group of Picture (GoP) per sequence thus expanding the utility of the dataset. As can be seen from Fig. 5 (a), where the mean and per GoP curves are plotted, using the per GoP data can help cover a wider range of curves. This is useful both for analysis and regression purposes. Furthermore, in order to assess how much the curves within a GoP might vary, we computed the Bjøntegaard Delta PSNR (BDPSNR)(41) of the per GoP curves over the mean. In Fig. 5 (b), we present the cumulative histogram of BDPSNR values of the RD curves per GOP over the RD of each sequence. As can be observed, in many cases the RD curves per GOP are very close in terms of quality to the overall RD curve of the sequence, evidenced by the mean value of the BDPSNR histogram being less than 0.01dB. However, the standard deviation of 1.81dB indicates a significant difference in the vertical shift of the RD curves. The difference between the mean and per GoP RDs becomes more significant when we inspect the difference in terms of bit rate in Fig. 5 (c), where the mean BDRate is 16.41% and the standard deviation is 61.39%. Finally, to better understand how the BD metrics are distributed in terms of texture type, we scattered BDPSNR against BDRate, coloured according to expert annotations in Fig. 5 (d). As can be seen, for most of the static sequences, although the BDPSNR is close to zero, the distribution of BDRates is very wide. On the other hand, it is noticeable that the variation of BDRate and BDPSNR demonstrates a similar curve for the case of dynamic textures.

### 3.2. RD Curves per Video Texture Type

In Fig. 6 (a), we illustrate the RD curves for all HomTex sequences coloured according to the expert annotations (red for static, blue for dynamic continuous and green for dynamic discrete textures) and in (b) the average RD curves per different texture type. As can be seen, from Fig. 6 (a), the sequences are generally naturally clustered. There are, however, cases, where certain sequences that were by context classified by the experts as one type of texture, behave similarly as another type of texture. In Fig. 6 (b), the vertical ranges denote the standard deviation of the PSNR, while the horizontal ranges denote the bit rate standard deviation. This figure shows the average RD performance per texture type and confirms with the overlapping standard deviation ranges that there are indeed sequences that perform similarly to other texture types. For example, according to experts, CalmSea is annotated as a dynamic continuous texture sequence. True to its name, the scene depicts sea water that is not moving fast. Therefore, it is easy to predict and thus compress.

### 3.3. Modelling the RD Curves
#### 3.3.1. RD Models

First, using ordinary least squares, we explore several different families of models, such as polynomial, exponential and power models to find the best fit for the log(R)-PSNR curves for video textures. The linear model was proposed in our previous work (20) and will serve as a basis for comparison. The $3^{rd}$-order polynomial is generally used to compute BD metrics (41) and will be explored here. In Table 3 the explored models are reported along with their goodness of fit values, particularly the Root MSE (RMSE) and the coefficient of determination $R^2$. It is clear from this Table that the best two candidates models are the $3^{rd}$-order polynomial (Poly3) and the $2^{nd}$-order polynomial (Poly2) model. Both have low RMSE values and very high $R^2 (> 0.99)$. Poly2 has the advantage of having only three parameters that need to be computed, while Poly3 has four parameters. This means that the training and prediction of the RD parameters with machine learning methods (Section 4) will require more computations.

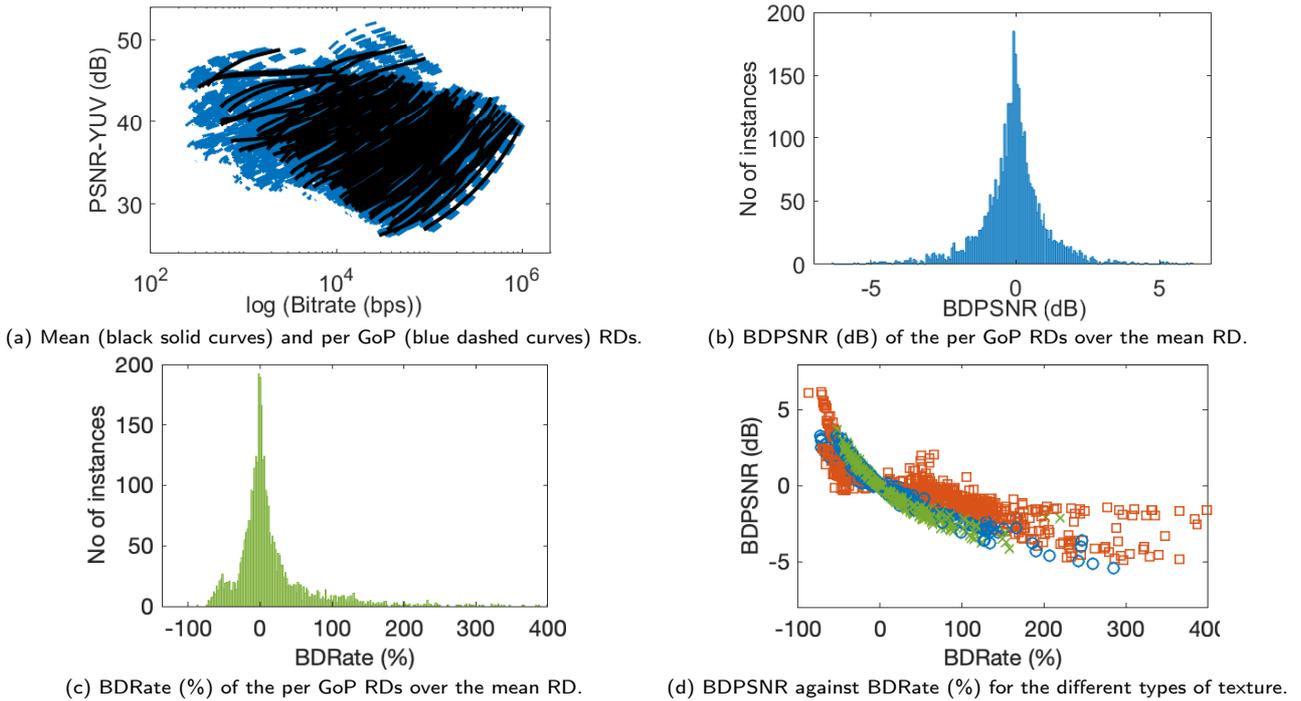

**Figure 5:** Comparison of the mean and per GoP RD curves for the HomTex dataset. The colours in subfigure (d) are used as follows: red for static, blue for dynamic continuous and green for dynamic discrete.

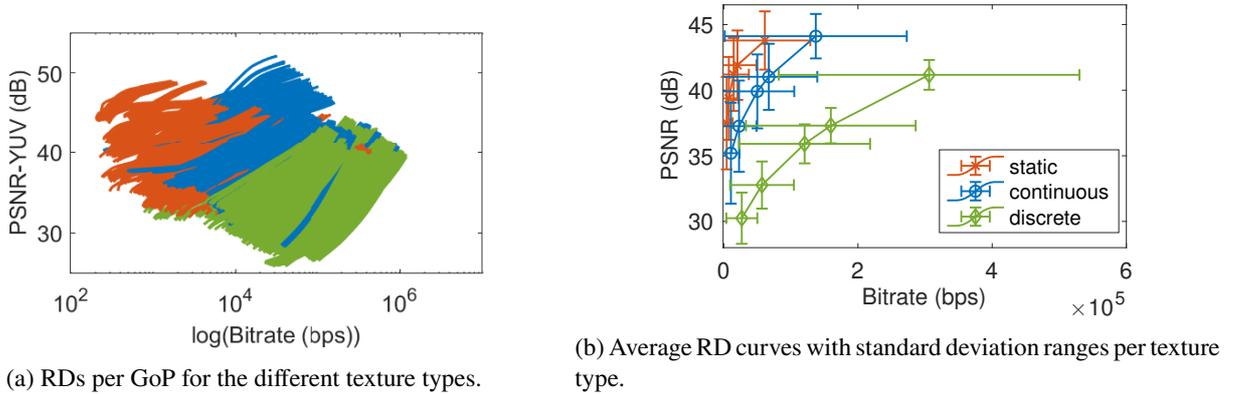

**Figure 6:** HomTex RDs per GoP and average RD curves for the different types of video texture. The different colours correspond to the sequence annotation by the experts: red for static, blue for dynamic continuous and green for dynamic discrete textures.

In the above context, we need to understand the tradeoff in terms of RD fitting performance, if one of the models with three parameters is selected. To select the most appropriate model we have computed the BD delta metrics between each fitted RD curve and its real version. The mean and standard deviation (std) BDPSNR and BDRate values are reported in the last two columns of Table 3. The BD metrics confirm that Poly3 and Poly2 are the two best performing models, with lower absolute mean values and stds. The performance of Poly2 and Poly3 are very similar in terms of BDPSNR. In terms of BDRate, Poly2 seems to be performing better on average. Although the Poly3 model does not demonstrate the lowest absolute mean BDRate value, it has the lowest std value compared to the other two. In terms of BDPSNR, Poly3 has both the lowest mean and std value. Regarding the other two models, Lin and Exp, they have a similar performance, with Exp achieving better BDPSNR statistics and Lin better BDRate statistics.

As the goodness of fit and BD metrics are very high and low, respectively, for the explored models and occupy a similar range, we further investigated the significance of their difference. We computed the empirical Cumulative

**Table 3**

Explored mathematical models for RD curves (where $\alpha_i, \beta_i, \gamma_i, \delta_i \in \mathbb{R}$, with $i \in \{1, \ldots, 4\}$) and goodness of fit assessment.

| Model | Acro. | Formula | $R^2$ | RMSE |
|---|---|---|---|---|
| Linear | Lin | $Q(R) = \alpha_1 \log(R) + \beta_1$ | .96 | .5477 |
| 2$^{\text{nd}}$-order Polynomial | Poly2 | $Q(R) = \alpha_2 \log^2(R) + \beta_2 \log(R) + \gamma_2$ | .99 | .1247 |
| 3$^{\text{rd}}$-order Polynomial | Poly3 | $Q(R) = \alpha_3 \log^3(R) + \beta_3 \log^2(R) + \gamma_3 \log(R) + \delta_3$ | .99 | .0402 |
| Exponential | Exp | $Q(R) = \alpha_4 e^{\beta_4 \log(R)}$ | .96 | .4300 |

**Table 4**

BD metrics for the explored mathematical models for RD curves. The $\mu \pm \sigma$ represent the mean value and the standard deviation.

| Model | BDPSNR: $\mu \pm \sigma$ | BDRate: $\mu \pm \sigma$ |
|---|---|---|
| Lin | -.0885 ± .1552dB | -.0075 ± 1.0416% |
| Poly2 | -.0094±.1274dB | -.0037±.2985% |
| Poly3 | -.0004±.0028dB | .0043±.0542% |
| Exp | -.0364±.1302dB | -.1534±1.0258% |

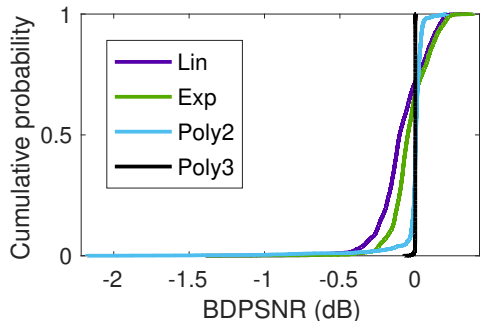
(a) CDF distributions of BDPSNR for the explored models.

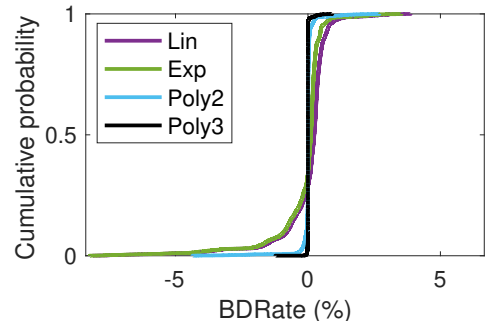
(b) CDF distributions of BDRate for the explored models.

**Figure 7:** Comparison of the performance of the fitted RD models.

Distribution Functions (CDF) and the histograms of the BD metrics and plot these in Fig. 7. As can be seen in Fig. 7 (a) and (b), Poly3 has the best performance as its variance around 0 is almost negligible. We note that Poly2 has a very similar CDF curve compared to Poly3 for both BDPSNR and BDRate. Lastly, the CDF curves for Lin and Exp are very similar.

From the discussion above, we conclude that Poly3 best fits the video texture RD curves. In order, however, to make a recommendation of which model to use for the prediction of the RD curves based on spatio-temporal features, we need to make a comparative analysis of the accuracy of the predicted RDs against the real curves for each tested model, as well as their relative complexity.

### 3.3.2. RD Model Parameters

In this Section, we study the RD curve model parameters and inspect their correlation to the characteristic shapes and parameters of the RD curves. For example, for the Poly model in Eq. (**??**), parameter $\alpha$ is related to the curvature of the RD curve. Parameter $\beta$ is related with the slope of the RD curve, which expresses the "cost" in bit rate in the logarithmic domain for a given increase video quality. Last, parameter $\gamma$ is related to the horizontal shift of the RD curve which approximates the minimum bit budget required for the video encoding.

From the RD curves plotted in Fig. 5, we can observe that the curvature is generally positive for lower bit rates, while it becomes negative for higher bit rates. This is confirmed by the Poly2 parameter $\alpha$ and its correlation to parameter $\gamma$; the Pearson Correlation Coefficient (PCC) is .9706 and the Spearman Rank Correlation Coefficient (SROCC) is .9867. A similar observation is made for parameters $\alpha$ and $\beta$. Particularly, when the slope is flatter usually the RD curve has a positive curvature and the inverse is observed for RD curves with steeper slopes. This is confirmed in Fig. 8 (a), where parameter $\alpha$ is scattered against $\beta$, and also by computing the correlation coefficients between the two parameters: PCC is equal to -.9939 and SROCC is equal to -.9979. Regarding the other parameters of Poly2, we notice in Fig. 5 that in general, the smaller the slope is, the more shifted towards lower bit rates the curve on the log($R$) axis is. This indicates a strong and potentially linear relation among those two parameters that is confirmed in Fig. 8 (b), where the parameter $\gamma$ is scattered against $\beta$. The PCC is very high, -.9908, as well as the SROCC, -.9944. Similar and correlations can be observed and conclusions drawn for the parameters of the other considered RD models.

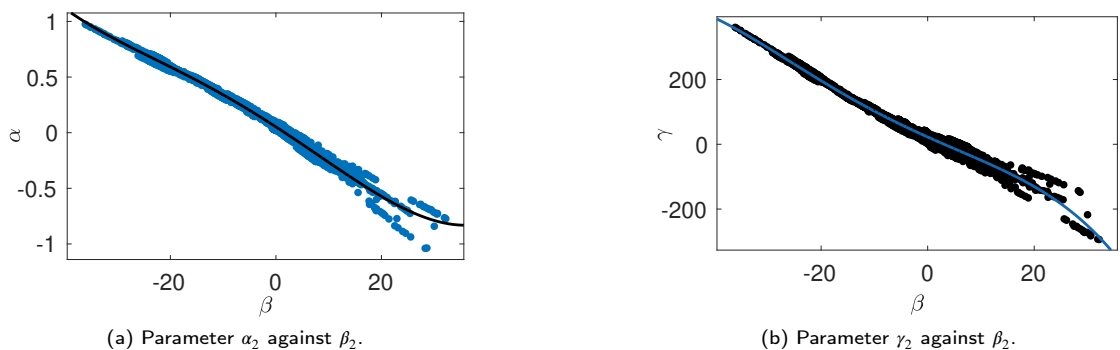

**Figure 8:** Example of RD model parameters relation: Poly2 parameters.

**Table 5**
Relation of the fitted RD model parameters.

| Model | RD Parameter Models | PCC | SROCC | $R^2$ |
|---|---|---|---|---|
| Lin | $\tilde{\beta}_1 = .8571\alpha_1^3 - 6.796\alpha_1^2 - 8.117\alpha_1 + 40.95$ | -.9825 | -.9846 | .97 |
| Poly2 | $\tilde{\alpha}_2 = 1.43 \cdot 10^{-7}\beta_2^4 + 4.22 \cdot 10^{-6}\beta_2^3 - 1.64 \cdot 10^{-4}\beta_2^2 - 3.08 \cdot 10^{-2}\beta_2 + 5.21 \cdot 10^{-2}$ | -.9939 | -.9979 | .99 |
|  | $\tilde{\gamma}_2 = -4.94 \cdot 10^{-5}\beta_2^4 - 1.97 \cdot 10^{-3}\beta_2^3 + 4.56 \cdot 10^{-2}\beta_2^2 - 7.38\beta_2 + 22.53$ | -.9908 | -.9944 | .99 |
| Poly3 | $\tilde{\alpha}_3 = -1.14 \cdot 10^{-9}\beta_3^5 - 2.01 \cdot 10^{-7}\beta_3^4 - 7.63 \cdot 10^{-6}\beta_3^3 - 1.57 \cdot 10^{-4}\beta_3^2 - .02\beta_3 + 1.59 \cdot 10{-3}$ | -.9829 | -.9973 | .99 |
|  | $\tilde{\delta}_3 = 8.26 \cdot 10^{-12}\gamma_3^5 - 1.94 \cdot 10^{-8}\gamma_3^4 + 1.03 \cdot 10^{-5}\gamma_3^3 + 2.53 \cdot 10^{-3}\gamma_3^2 - 6.21\gamma_3 + 26.64$ | -.9840 | -.9968 | .99 |
| Exp | $\tilde{\beta}_4 = -.551\alpha_4^{.064} + .711$ | -.9440 | -.9919 | .98 |

The linear and rank correlation values of the RD parameter models are reported in Table 5. In order to take advantage from the high correlations of the RD model parameters, we explored several mathematical models and the ones that achieved the highest goodness of fit values (e.g. $R^2$) are given in Table 5. Modelling the RD parameters provides the benefit of training fewer models, only for the parameters that are required to estimated the other ones. For example, for the Lin model, we can build a machine-learning based regression model to predict parameter $\beta_1$ and then with Eqs.(??)-(??) we can estimate the other parameter $\alpha_1$.

## 4. Predicting RD Curves using Textural Features from Uncompressed Content

Previous work that relates textural features to video compression efficiency includes (42; 43; 30). In (30), Subedar et al. define a no-reference metric of granularity in static textured images and discuss its relation to compression efficiency, but with no clear association with RD curves. In (43), elementary statistics of prediction error for texture and motion are obtained from H.264/AVC encoder and used to build variability-distortion models. In (42), a block-based spatial correlation model is defined and used to predict the RD bounds within an H.264/AVC encoder. This work was extended in (44) to consider the block-based spatial correlation among two successive frames within HEVC HM. In our work, we predict RD curves for homogeneous video textures by building models driven by spatio-temporal features extracted from the uncompressed source sequences. We consider RD parameter correlations and reduce the prediction complexity by using the models explored in Section 3. This extends our previous work (20), through a full characterisation and comparison of RD models (previous Section), by exploring machine learning based regression methods and by recommending the RD model taking into account the complexity-accuracy tradeoff.

**Table 6**
Validation metrics of predicted parameters and the resulting BD metrics for the tested RD models. The $\mu \pm \sigma$ represent the mean value and the standard deviation.

| Model | Param. | Features/Eqs. | PCC | SROCC | $R^2$ | MAE | NRMSE |
|---|---|---|---|---|---|---|---|
| **Lin** | $\hat{\alpha}_1$ | F1, F4, F15, F22, F24, F26, F28, F30, F32-F35, F37 | .9303 | .9247 | .86 | .2173 | .0817 |
| | $\hat{\beta}_1$ | $\tilde{\beta}_1$, Table 5 | .9215 | .9182 | .85 | 5.9199 | .0886 |
| **Poly2** | $\hat{\alpha}_2$ | $\tilde{\alpha}_2$, Table 5 | .9311 | .9475 | .87 | .0632 | .0521 |
| | $\hat{\beta}_2$ | F1, F4, F15, F22, F24, F26, F28, F30, F32-F35, F37 | .9327 | .9481 | .87 | 2.1676 | .0557 |
| | $\hat{\gamma}_2$ | $\tilde{\gamma}_2$, Table 5 | .9206 | .9420 | .85 | 19.7760 | .0544 |
| **Poly3** | $\hat{\alpha}_3$ | F1, F3, F4, F15, F22, F24, F26, F28, F30, F32-F35 | .6508 | .7922 | .44 | 0.0273 | .0224 |
| | $\hat{\beta}_3$ | F1, F4, F15, F22, F24, F26, F28, F30, F32, F33 | .6359 | .7982 | .53 | 1.4056 | .0227 |
| | $\hat{\gamma}_3$ | F1, F4, F5, F15, F22, F24, F26, F28, F30, F32-F35, F37 | .7518 | .8077 | .58 | 23.6200 | .0277 |
| | $\hat{\delta}_3$ | $\tilde{\delta}_3$, Table 5 | .7884 | .8083 | .62 | 141.1246 | .0332 |
| **Exp** | $\hat{\alpha}_4$ | F1, F4, F15, F22, F24, F26, F28, F30, F32-F35, F37 | .9440 | .9346 | .89 | 1.7671 | .0816 |
| | $\hat{\beta}_4$ | $\tilde{\beta}_4$, Table 5 | .9124 | .9228 | .83 | .0077 | .1062 |

### 4.1. Experimental Setup

For the training and testing of the proposed method, the first 240 frames of the HomTex (7) sequences were used (some of the sequences have 250 frames and 25 fps frame rate and others 300 frames and 60 fps frame rate). The RD curves were obtained using the HM16.20 reference software in Random Access configuration for five different quantization scales, QP = {22, 27, 32, 37, 42}, GoP length equal to 8 and Intra Period 32. The RD curves were constructed per GoP resulting in 3600 RD curves. For all RD curves, we fitted the RD models of Table 3 and computed their parameters as explained in Section 3.3.2. These fitted parameters serve as the ground truth for the regression tasks. Furthermore, we extracted the spatio-temporal features from the uncompressed source sequences and computed the statistics per GoP. Finally, we normalised the features.

Before building and testing regression models for all considered RD models, we perform feature selection in order to reduce the computational cost of feature extraction. We have noticed before (21) that many of the features are correlated, thus different combination of features may achieve the same prediction accuracy. We perform a feature selection and elimination process based on RF models, the Recursive Feature Elimination (RFE) (45). RFE uses feature ranking and iteratively selects subsets of features with different cardinality and returns the optimal feature subset. We note that we perform feature selection for each one of the predicted RD parameters, as the subset of selected features varies slightly.

To avoid overfitting, a random 10-fold cross-validation process was adopted. A random split of the extracted features and the fitted RD parameters was performed with 80% of the data being used for model configuration and training and the remaining 20% of the data for the performance evaluation. The machine-learning regression models explored included: Support Vector Machines (SVMs), Random Forests (RFs), Ensemble Trees and Gaussian Processes (GPs) using different kernels and by optimising the associated parameters.

### 4.2. Results

The evaluation of the performance of the proposed approach takes place in two steps. First, the accuracy of predicting RD parameters is assessed. Next, the predicted RD curves are validated against the real ones using BD metrics. Finally, the results are discussed in terms of the accuracy of prediction and their relative complexity.

**Table 7**
Resulting BD metrics for the tested RD models. The $\mu \pm \sigma$ represent the mean value and the standard deviation.

| Model | BDPSNR: $\mu \pm \sigma$ | BDRate: $\mu \pm \sigma$ | RCR |
|---|---|---|---|
| **Lin** | .0973±4.5434dB | -.7379±14.1567% | 1 |
| **Poly2** | .3811±3.1146dB | -3.6770±17.2721% | 1.21 |
| **Poly3** | 2.2904±51.0300dB | -32.2441±58.0606% | 3.14 |
| **Exp** | -.0958±4.0885dB | **.4552±11.6676%** | 1.08 |

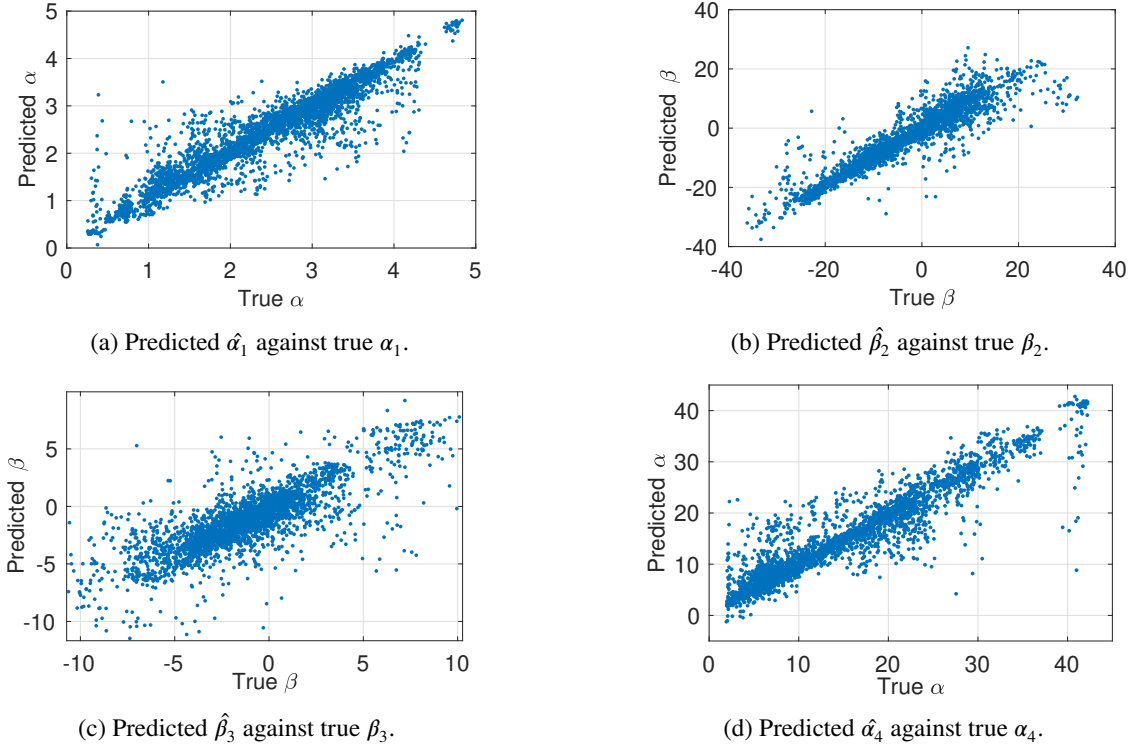

**Figure 9:** Examples of predicted RD parameters against the true (fitted) values for the four different RD models.

*4.2.1. Accuracy of RD parameters prediction*

In Table 6, a summary of the best results achieved per different RD model are presented. Particularly, for each predicted RD parameter the following evaluation metrics are reported:

- the selected features or the equation (if that produced better results);
- the correlation metrics, PCC and SROCC between the predicted and the fitted RD parameter;
- the coefficient of determination $R^2$, the Mean Absolute Error (MAE), and the NRMSE.

It is remarkable that although Poly3 was the best fitted model (see Section 3.3.1), the prediction of the parameters of the lower order models, Lin, Exp and Poly2, is of higher accuracy as the correlation metrics are greater than .9 and the coefficient of determination is higher than .83. The improved prediction of the RD parameters for the lower order models is noticeable also in the example plots in Fig. 9. In this figure, we illustrate examples of the predicted RD parameter values against their true (fitted) values for the different tested models. As can be seen, the distribution around the main diagonal is tighter for Lin, Poly2 and Exp RD models explaining the high validation metric values in Table 6. On the other hand, Fig. 9 (c) justifies the low PCC and SROCC values for $\beta_3$ parameter as the predicted vs the true values are not tightly distributed around the diagonal.

*4.2.2. Accuracy of RD curve prediction*

To fully validate the effectiveness of the proposed method, we computed and report in Table 6:

- the mean and standard deviation of the BDPSNR (dB) and BDRate (%) of the predicted RD curves over the real ones;

- the Relative Complexity Ratio (RCR), i.e. the ratio of execution time for each model over the minimum execution time recorded.

As anticipated from the validation metric values associated with the predicted parameters, the mean BDPSNR and BDRate are higher for Poly3. The predicted curves are closer to the real ones in terms of average and standard deviation of BDRate for the Exp model. Exp and Lin have very similar mean BDPSNR values, however Exp has a lower standard deviation. Poly2 has the lowest standard deviation value, but bit a higher mean value.

Regarding the complexity, this depends on the number of features that need to be extracted (dictated by the offline feature selection process) and on the number of machine learning-based predictions that need to be performed. For some models, it has been shown that the fitted equations can be reliable for an accurate RD parameter estimation. The low order models, benefit of having only two parameters that need to be determined and for one of them a fitted equation can be used. A lower number of predicted parameters is directly related to the decreased accumulated prediction error. Indeed, the models with only two parameters, Lin and Exp, demonstrate the lowest complexity and highest accuracy, while for the four-parametric model Poly3 the opposite is shown.

Taking into account all the above points, we can conclude that Poly3 is theoretically the best RD model. Nevertheless, the Exp model emerges as the prominent solution that balances accuracy and complexity, as it achieves the highest accuracy of prediction for the BDRate metric and the second lowest RCR amongst the compared models.

## 5. Conclusion

Predicated on the challenges of compressing video textures, we have characterised their spatio-temporal features and their encoding statistics using homogeneous video sequences. We have employed a homogeneous dataset with 120 sequences, HomTex. We have investigated mathematical models capable of representing the corresponding RD curves and characterised their accuracy. For all considered theoretical models, we employ machine-learning regression to predict the RD curves based exclusively on the selected spatio-temporal feature statistics that were extracted from uncompressed video content. Taking into account the tradeoff between the RD prediction accuracy and the relative computational complexity, we conclude that the exponential model performs best. The proposed method can be modified to effectively predict other encoding decisions or statistics, which remains part of our future work.